\documentclass[conference]{IEEEtran}
\IEEEoverridecommandlockouts

\usepackage[utf8]{inputenc}
\usepackage[T1]{fontenc}
\usepackage{cite}
\usepackage{svg}

\ifCLASSINFOpdf
\else
   \usepackage[dvips]{graphicx}
\fi
\usepackage{url}
\usepackage{graphicx}
\usepackage{ragged2e}
\usepackage{array} 
\usepackage{booktabs}
\usepackage{scalefnt}
\usepackage{graphics}
\usepackage{amssymb}
\usepackage{amsmath}
\usepackage{subeqnarray}
\usepackage{multirow}
\usepackage{sidecap}
\usepackage{lscape}
\usepackage{tabularx}
\usepackage{times}

\usepackage[ruled,vlined]{algorithm2e}
\usepackage{algorithmic}
\usepackage{listings,lstautogobble}
\usepackage{mathtools}

\DeclareMathOperator*{\argmin}{arg\,min}
\newcommand{\pinv}{\dag}
\newcommand{\ma}[1]{\boldsymbol{#1}}
\newcommand{\compl}{\mathbb{C}}

\newcommand{\eqdef}{\stackrel{.}{=}}
\newcommand{\diagof}[1]{{\rm diag}\left\{ #1 \right\}}

\newcommand{\ten}[1]{\mathcal #1}
\newcommand{\fronorm}[1]{\left|\left|#1\right|\right|_{\rm F}}

\newcommand{\krp}{\diamond}

\newcommand{\trans}{{{\rm T}}}
\newcommand{\herm}{{{\rm H}}}

\newcommand{\mat}[1]{\ensuremath{{\mathbf{#1}}}}
\newcommand{\vet}[1]{\mathit{\boldsymbol{#1}}}

\renewcommand{\baselinestretch}{0.99}

\begin{document}

\title{Channel Estimation for Downlink Communications Based on Dynamic Metasurface Antennas
\thanks{The authors acknowledge the partial support of the National Institute of Science and Technology (INCT-Signals) sponsored by Brazil's National Council for Scientific and Technological Development (CNPq) under grant 406517/2022-3. This work is also partially supported by CNPq under grants 312491/2020-4 and 443272/2023-9.}
}
\author{\IEEEauthorblockN{Amarilton L. Magalhães\IEEEauthorrefmark{1}\IEEEauthorrefmark{3}, Fazal E-Asim\IEEEauthorrefmark{1}, André L. F. de Almeida\IEEEauthorrefmark{1} and A. Lee Swindlehurst\IEEEauthorrefmark{2}}\\
\IEEEauthorblockA{Federal University of Ceará, Brazil\IEEEauthorrefmark{1}\\Federal Institute of Education, Science and Technology of Ceará, Brazil\IEEEauthorrefmark{3}\\University of California at Irvine, CA, USA\IEEEauthorrefmark{2}
\\E-mail: amarilton@gtel.ufc.br, fazalasim@gtel.ufc.br, andre@gtel.ufc.br, swindle@uci.edu}
}
\maketitle

\begin{abstract}
Dynamic metasurface antennas (DMAs) are emerging as a promising technology to enable energy-efficient, large array-based multi-antenna systems. This paper presents a simple channel estimation scheme for the downlink of a multiple-input single-output orthogonal frequency division multiplexing (MISO-OFDM) communication system exploiting DMAs. The proposed scheme extracts separate estimates of the wireless channel and the unknown waveguide propagation vector using a simple iterative algorithm based on the parallel factor (PARAFAC) decomposition. Obtaining decoupled estimates of the wireless channel and inner waveguide vector enables the isolation and compensation for its effect when designing the DMA beamformer, regardless of the wireless channel state, which evolves much faster due to its shorter coherence time and bandwidth. Additionally, our solution operates in a data-aided manner, delivering estimates of useful data symbols jointly with channel estimates, without requiring sequential pilot and data stages. To the best of our knowledge, this is the first work to explore this CE approach. Numerical results corroborate the notable performance of the proposed scheme.
\end{abstract}


\IEEEpeerreviewmaketitle

\section{Introduction}\label{sec:introduction}
Dynamic metasurface antennas (DMAs) have attracted increasing interest in wireless communications as a promising technology for next-generation networks, such as 6G \cite{shlezinger2021dynamic}. DMAs can enable planar, low-cost, energy-efficient, extra-large antenna arrays, since their construction facilitates hybrid analog/digital precoding, by integrating dynamic analog combining capabilities directly into the antenna structure, in contrast to conventional multiple-input multiple-output (MIMO) technologies, which rely on fixed array designs and complex signal processing \cite{shlezinger2019dynamic, shlezinger2021dynamic, carlson2023dynamic, rezvani2024channel}. Channel estimation (CE) is a crucial and challenging task in metasurface-based wireless communication systems, such as reconfigurable intelligent surfaces (RISs) \cite{swindlehurst2022channel, basar2024reconfigurable}, being mandatory for designing passive beamforming as well as precoder/combiner at the BS/UE. For DMA-based systems, the CE literature is still in its embryonic stage, except for the pioneering article attributed to \cite{rezvani2024channel}, where a CE method utilizing pilot sequences and leveraging the fast switching of DMA elements was proposed. However, the authors relied on the knowledge of the channel statistics and assumed perfect knowledge of the electromagnetic field vector, which models the signal propagation inside the DMA microstrips. In a practical scenario, exact knowledge of this vector may not be available due to the non-homogeneous behavior of the radiating elements or mismatches between the idealized waveguide model and the actual construction of the DMA.

Many papers in the wireless communication literature have addressed problems using tensor decompositions \cite{Almeida2016, Miron2020}, applying them to model blind/semi-blind receivers \cite{sidiropoulos2000blind}, space-time (ST) \cite{favier2012tensor}, space-time-frequency (STF) coding \cite{favier2014tensor} schemes, and relay-assisted MIMO systems \cite{ximenes2014parafacparatuck}. Recent works have also exploited tensor tecompositions in the context of programmable metasurfaces, such as channel estimation (CE) \cite{dearaujo2021channel}, \cite{gomes2023channel}, semi-blind joint CE and symbol detection \cite{magalhaes2025reducing}, feedback design \cite{Sokal_2023}, and channel tracking \cite{benicio2024tensor_wcl}.

In this work, we refer to this waveguide propagation vector as the DMA inner channel vector. In this case, teasing out a separate estimation of the outer wireless and the inner channels allows us to isolate the two effects and adapt the beamforming design accordingly. Additionally, integrating CE with data detection enables reuse of the actual data symbols to enhance channel estimates while minimizing symbol decoding delays \cite{dearaujo2023semiblind}. Previous work has not addressed these issues. We propose a simple CE scheme for an MISO-OFDM system assisted by DMAs. By exploiting fast time-switching beamformers, as noted by \cite{rezvani2024channel}, our method employs a PARAFAC-based algorithm to provide direct estimates of the data symbols and utilizes them to iteratively refine the channel estimates, thereby eliminating the need for separate pilot training and pure data transmission phases. Simulations demonstrate accurate estimation under unknown waveguide conditions and reveal the influence of the parameters on performance.

\subsection{Notations}
We utilize $a$, $\vet{a}$, $\mat{A}$, and $\ten{A}$ for scalars, vectors, matrices, and tensors, respectively. A tensor $\ten{X}\! \in \!\compl^{I \times J \times \cdots \times I_P}$ is a $N$-dimensional array. The operations $\mat{A}^\trans$, $\mat{A}^\herm$, and $\mat{A}^\pinv$ stand for the transpose, Hermitian, and Moore-Penrose pseudo-inverse of $\mat{A}$, respectively. The Frobenius norm is indicated by $\fronorm{\cdot}$, and the Hadamard, Khatri-Rao, and outer matrix products are denoted by $\odot$, $\krp$, and $\circ$, respectively. In addition, $\diagof{\vet{a}}$ forms a diagonal matrix from $\vet{a}$, and $D_k(\mat{A})$ returns a diagonal matrix constructed from the $k$-th row of $\mat{A}$.

\subsection{PARAFAC decomposition}
The PARAllel FACtors decomposition was originally proposed by \cite{carroll1970analysis} and \cite{harshman1970foundations} independently, and applied in the research field of wireless communications for the first time in \cite{sidiropoulos2000blind}. The principle of the PARAFAC decomposition is to factorize a tensor into a sum of rank-one tensors. For example, a third-order tensor $\ten{X} \!\in\!\compl^{I \times J \times K}$ can be expressed in the scalar form as $x_{i,j,k} \!=\! \textstyle \sum_{n=1}^N{a_{i,n}b_{j,n}c_{k,n}}$ or, in the tensor form as $\ten{X} \!=\! \textstyle \sum_{n=1}^N{\vet{a}_n \!\circ\! \vet{b}_n \!\circ\! \vet{c}_n}$, where $(a_{i,n}, b_{j,n}, c_{k,n})$ and ($\vet{a}_n \!\!\in\!\! \compl^{I \times 1}\!, \vet{b}_n \!\!\in\!\! \compl^{J \times 1}\!, \vet{c}_n \!\!\in\!\! \compl^{K \times 1}\!$) are, respectively, the scalar components and columns of the factor matrices $(\mat{A} \!\!\in\!\! \compl^{I \times N}\!, \mat{B} \!\in\! \compl^{J \times N}\!, \mat{C} \!\in\! \compl^{K \times N})$. The PARAFAC decomposition of $\ten{X}$ can also be expressed in a more modern, compact form by means of the $n$-mode product formalism \cite{kolda2009tensor}, given by
\begin{equation}\label{parafacnmodeprod}
    \ten{X} = \ten{I}_{3,N} \times_1 \mat{A} \times_2 \mat{B} \times_3 \mat{C},
\end{equation}
where $\ten{I}_{3,N}$ denotes the identity tensor, while its $k$-th frontal slice representation is given by
\begin{equation}\label{parafacfrontalslice}
    \mat{X}_k = \mat{A}D_k(\mat{C})\mat{B}^\trans.
\end{equation}

\section{System model and assumptions}\label{sec:system}
\begin{figure}[!t]
	\centering
	\includegraphics[width=0.43\textwidth]{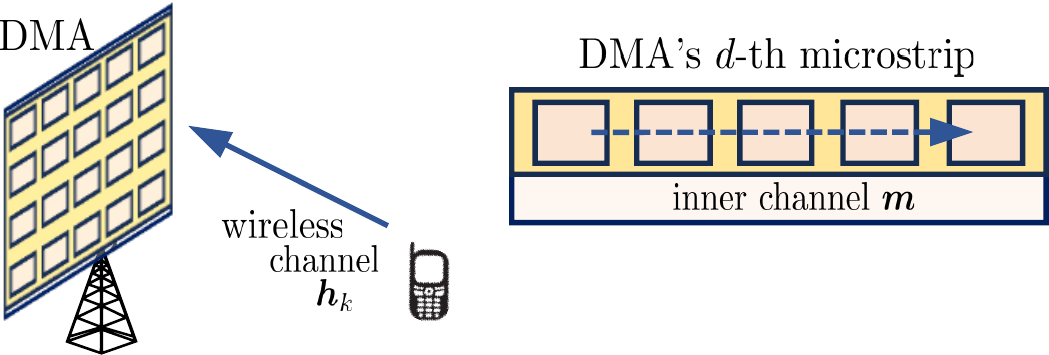}
    \caption{Illustration of the adopted system.}
	\label{fig:system}
\end{figure}

We consider a downlink multiple-input single-output (MISO) wireless communication system based on orthogonal frequency division multiplexing (OFDM) using DMA at the transmit side (base station), depicted in Figure \ref{fig:system}. The DMA is comprised of $D$ waveguides, each with $L$ radiating elements, resulting in a total of $N = DL$ elements. The transmitted signal comprises $K$ subcarriers. The DMA inner channel, i.e., signal propagation across the waveguides, is modeled by $\vet{m} \eqdef [m_1, \ldots, m_N]^\trans \in \compl^{N \times 1}$, with $m_n = m_{d,l} = e^{-(\alpha+ j\beta)x_{d,l}}$, where $x_{d,l}$ defines the position of the $l$-th element of the $d$-th waveguide, and $\alpha$ and $\beta$ are the waveguide attenuation and propagation constants, respectively. We adopt a narrowband model where $\alpha$ and $\beta$ are approximately unchanged inside a narrow bandwidth, such that $\ma{m}$ is assumed to be frequency-independent \cite{carlson2023dynamic,rezvani2024channel}.We define $\vet{h}_k \eqdef [h_{1,k}, \ldots h_{N,k}]^\trans \in \compl^{N \times 1}$ as the wireless channel vector related to the $k$-th subcarrier. Defining $\vet{f}^{\text{DMA}} \in \compl^{N \times 1}$ as the DMA beamformer and $s$ as the transmitted symbol reused across the $K$ subcarriers, we can write the received signal as 
\begin{equation}\label{eq:rxsig}
	y_k = \vet{h}^{\herm}_{\text{eff},k}\vet{f}^\text{DMA}s + n_k,
\end{equation}
where $\vet{h}_{\text{eff},k} = \vet{h}_k^\herm \odot \vet{m} \in \compl^{N \times 1}$ is the \textit{effective channel vector} by incorporating the inner channel into the wireless one, and $n_k$ is the additive white Gaussian noise. Following \cite{carlson2023dynamic}, we assume that the DMA beamformer has a frequency-flat response over a narrow bandwidth, acting as a bandpass filter, whose amplitude and phase result from the meta-atom's magnetic polarizability at a specific varactor diode tuning state, satisfies the so-called Lorentzian constraints, i.e., $Q = \{\frac{j+ e^{j\phi}}{2} | \phi \in [0, 2\pi] \}$ \cite{smith2017analysis, shlezinger2019dynamic}.

We propose a CE scheme that (i) yields separate estimates of these two channel components and (ii) incorporates data detection still during the CE phase. Identifying the inner channel component allows us to isolate and compensate for its effect when designing the DMA beamformer, regardless of the wireless channel state, which evolves much faster due to its shorter coherence time and bandwidth. We consider a transmission window of $T$ symbol periods, where the DMA beamforming vector, denoted by $\vet{f}^\text{DMA}_p$ for the $p$-th state, switches $P$ times per symbol by leveraging the fast response of DMAs \cite{rezvani2024channel}. The received signal associated with the $k$-th subcarrier, $t$-th symbol period, and $p$-th beamforming state can be written as $y_{k,t,p} = \vet{h}^{\herm}_{\text{eff},k}\ma{f}^\text{DMA}_ps_t + n_{k,t,p}$, or $y_{k,t,p} = \vet{h}^{\herm}_k\diagof{\vet{m}}\vet{f}^\text{DMA}_ps_t + n_{k,t,p}$, by decoupling $\vet{h}_k$ and $\vet{m}$. Using the property $\diagof{\vet{a}}\vet{b} = \diagof{\vet{b}}\vet{a}$ to exchange $\vet{m}$ and $\ma{f}^\text{DMA}_p$, and collecting the received data across $T$ symbols and $K$ subcarriers yields
\begin{equation}\label{matyp1}
	\mat{Y}_p = \mat{H}\mathrm{diag}\bigl\{\vet{f}^\text{DMA}_p\bigr\}\vet{m}\vet{s}^\trans + \mat{N}_p \in \compl^{K \times T},
\end{equation}
where $\vet{s} = [s_1,\ldots,s_T]^\trans \in \compl^{T \times 1}$, $\mat{H} = [\vet{h}_1, \ldots, \vet{h}_N]^\herm \in \compl^{K\times N}$ is the frequency-domain MISO channel matrix, and $\mat{N}_p$ is the related noise term. Defining $\mat{F} = [\vet{f}^\text{DMA}_1\!\!\!\!\!,\ldots,\vet{f}^\text{DMA}_P]^\trans \in \compl^{P \times N}$ as the DMA training matrix collecting the beamforming vectors for the $P$ states, and the rank-1 matrix $\mat{X} = \vet{s}\vet{m}^\trans \in \compl^{T \times N}$, we recast \eqref{matyp1} as
\begin{equation}\label{matyp2}
	\mat{Y}_p = \mat{H}D_p(\mat{F})\mat{X}^\trans + \mat{N}_p.
\end{equation}

\section{Tensor modeling}
According \eqref{parafacnmodeprod} and \eqref{parafacfrontalslice}, the received signal in \eqref{matyp2} can be interpreted as the $p$-th frontal matrix slice of a three-way tensor $\ten{Y} \in \compl^{K \times T \times P}$, whose signal component admits a parallel factor (PARAFAC) decomposition \cite{kolda2009tensor}, which can be written in tensor form as
\begin{equation}\label{teny}
	\ten{Y} = \ten{I}_{3,N} \times_1 \mat{H} \times_2 \mat{X} \times_3 \mat{F} + \ten{N},
\end{equation}
where $\ten{N}$ is the associated noise tensor. According to \cite{kolda2009tensor}, the mode-1 and mode-1 matrix unfoldings of $\ten{Y} \in \compl^{K \times T \times P}$ are $\mat{Y}_{(1)} \eqdef [\mat{Y}_1,\ldots, \mat{Y}_P] \in \compl^{K \times PT}$ and $\mat{Y}_{(2)} \eqdef [\mat{Y}_1^\trans,\ldots, \mat{Y}_P^\trans] \in \mathbb{C}^{T \times PK}$, which result in $\mat{Y}_{(1)} = \mat{H}(\mat{F} \diamond \mat{X})^\trans + \mat{N}_{(1)}$ and $\mat{Y}_{(2)} = \mat{X}(\mat{F}\diamond \mat{H})^\trans + \mat{N}_{(2)}$, respectively, where $\mat{N}_{(1)}$ and $\mat{N}_{(2)}$ are unfoldings of $\ten{N}$.

A starting condition to analyze the uniqueness of the PARAFAC decomposition is Kruskal's condition \cite{kruskal1977three, sidiropoulos2000uniqueness}. Bringing this concept to this work, by taking the factor matrices in \eqref{teny}, the tensor $\ten{Y}$ would be unique if $\kappa_{\mat{H}} + \kappa_{\mat{X}} + \kappa_{\mat{F}} \geq 2R + 2$, where $\kappa_{\mat{A}} \leq \mathrm{rank}(\mat{A})$ denotes the Kruskal-rank of $\mat{A}$ \cite{kruskal1977three}. Assuming that third factor matrix $\mat{F}$ is designed to have full column-rank, which implies $P \geq N$ and, taking into account $\mat{X}$ has rank one, the later inequality would be reduced to $\kappa_{\mat{H}} \geq N + 1$, which could not be satisfied. In fact, Kruskal's condition generally imposes a more restrictive constraint. 

The works \cite{jiang2004kruskal}, \cite{delathauwer2006alink}, \cite{stegeman2008uniqueness}, and \cite{guo2011candecomp} demonstrated necessary and sufficient uniqueness conditions more relaxed than the Kruskal one. Supposing at least one factor is full column-rank, for instance, the matrix $\mat{C}$, the authors in \cite{jiang2004kruskal} proved that if a specific matrix (see matrix $\mat{U}$ therein), that follows a particular construction based on the other two factors $\mat{A}$ and $\mat{B}$, is also full column-rank, then the PARAFAC decomposition is unique. In addition, \cite{delathauwer2006alink} shown that if the constraint $J(J-1)I(I-1)/4 \geq N(N-1)/2$ is stand, $\mat{U}$ has almost surely full column-rank when entries of $\mat{A}$ and $\mat{B}$ are randomly drawn from a continuous distribution of dimension $(I+J)N$. A structured explanation on the construction of ``$\mat{U}$'' is presented in \cite{stegeman2008uniqueness}. More contributions are found in \cite{delathauwer2006alink}. In our preliminary validations, by replacing $(\mat{A},\mat{B},\mat{C})$ by $(\mat{H},\mat{X},\mat{F})$, we have found ``$\mat{U}$'' full column-rank, even $\mat{X}$ being rank one. For this verification, we have considered that i) the MISO channel matrix $\mat{H}$ (designed as full column-rank) entries are taken from a zero-mean independent and identically distributed (i.i.d.) complex-valued Gaussian distribution; ii) the entries of the DMA propagation vector $\vet{m}$ are generated from exponentials whose arguments are randomly drawn from the interval $[0,2\pi)$; and iii) the transmitted symbols composing the vector $\vet{s}$ are generated from a QAM constellation.

\section{Data-aided semi-blind channel estimation}
Our first goal is to jointly estimate $\mat{H}$ and the rank-1 matrix $\mat{X}$. We consider solving the multilinear optimization problem
\begin{equation}\label{problem1}
	\min \limits_{\mat{H},\mat{X}} \Big\|\ten{Y} - \ten{I}_{3,N} \times_1 \mat{H} \times_2 \mat{X} \times_3 \mat{F}\Big\|^2_{\textrm{F}}.
\end{equation}
Problem \eqref{problem1} can be tackled by employing the iterative alternating least-squares (ALS) algorithm to estimate $\mat{H}$ and $\mat{X}$ by capitalizing on the mode-1 and mode-2 unfoldings of $\ten{Y}$. This is carried out by solving the problems
\begin{align}
	\hat{\mat{H}} &= \argmin_{\mat{H}}~\bigl\|\mat{Y}_{(1)} - \mat{H}(\mat{F} \diamond \mat{X})^\trans\bigr\|_\mathrm{F}^2,\\
    \hat{\mat{X}} &= \argmin_{\mat{X}}~\bigl\|\mat{Y}_{(2)} - \mat{X}(\mat{F} \diamond \mat{H})^\trans\bigr\|_\mathrm{F}^2.
\end{align}
The resulting solutions are given, respectively, by
\begin{align}
	\label{hest} \hat{\mat{H}} &= \mat{Y}_{(1)}\bigl[(\mat{F} \diamond \mat{X})^\trans\bigr]^\pinv,\\
    \label{xest} \hat{\mat{X}} &= \mat{Y}_{(2)}\bigl[(\mat{F} \diamond \mat{H})^\trans\bigr]^\pinv.
\end{align}
At each iteration, $\hat{\mat{H}}$ and $\hat{\mat{X}}$ are updated until convergence, which is declared when the change in reconstruction error $|\epsilon_{(i)} - \epsilon_{(i-1)}|$ becomes sufficiently small, where $\epsilon_{(i)} = \bigl|\ten{Y} - \hat{\ten{Y}}{(i)}\bigr|$, and $\hat{\ten{Y}}{(i)}$ denotes the reconstructed tensor at the $i$-th iteration. Once $\hat{\mat{H}}$ and $\hat{\mat{X}}$ are found, we consider recovering the symbol and inner channel vectors $\vet{s}$ and $\vet{m}$, respectively, from $\hat{\mat{X}}$. To address this problem, we solve the following rank-1 matrix approximation problem as a second stage:
\begin{equation}\label{problem2}
	\min \limits_{\vet{s},\vet{m}} \bigl\|\hat{\mat{X}} - \vet{s}\vet{m}^\trans\bigr\|^2_{\textrm{F}}.
\end{equation}
The best rank-1 approximation to solve \eqref{problem2} can be obtained via singular value decomposition (SVD), i.e., $\hat{\mat{X}} = \sigma\vet{u}\vet{v}^\herm$, whose optimal solutions are given by $\hat{\vet{s}} = \sqrt{\sigma}\vet{u}$ and $\hat{\vet{m}} = \sqrt{\sigma}\vet{v}^\ast$, respectively. The main procedures of the proposed two-stage CE method are provided in Algorithm \ref{alg:alg1}.

\begin{algorithm}[t]
 \small
 \caption{Two-stage semi-blind receiver}
 \label{alg:alg1}
    \vspace{1ex}
    \hspace{-3ex} \textit{Bilinear ALS-PARAFAC}\\
 \begin{algorithmic}
\STATE \hspace{-2ex} 1. Set $i=0$ and initialize $\hat{\mat{X}}_{(i=0)}$ randomly;\\
\STATE \hspace{-2ex} 2. $i = i + 1$;\\
\STATE \hspace{-2ex} 3. Get $\hat{\mat{H}}_{(i)} = \mat{Y}_{(1)}\bigl[(\mat{F} \diamond \mat{X}_{(i-1)})^\trans\bigr]^\pinv$;\\
\STATE \hspace{-2ex} 4. Get $\hat{\mat{X}}_{(i)} = \mat{Y}_{(2)}\bigl[(\mat{F} \diamond \mat{H}_{(i)})^\trans\bigr]^\pinv$;\\
\STATE \hspace{-2ex} 5. Repeat steps 2-5 until convergence;\\
\STATE \hspace{-2ex} 6. Remove scaling ambiguities.
 \end{algorithmic}
    \hspace{-3ex} \textit{Rank-1 approximation}\\
    \begin{algorithmic}
  \small{
   \STATE \hspace{-2ex} 7. Compute $[\vet{u},\sigma,\vet{v}] \longleftarrow$ truncated-SVD$(\hat{\mat{X}})$;\\
   \STATE \hspace{-2ex} 8. Estimate $\hat{\vet{s}}$ and $\hat{\vet{m}}$: ~~$\hat{\vet{s}} = \sqrt{\sigma}\vet{u}$,~~ $\hat{\vet{m}} = \sqrt{\sigma}\vet{v}^\ast$;\\
   \STATE \hspace{-2ex} 9. Remove scaling ambiguities.\\
    }
 \end{algorithmic}
\end{algorithm}

The overall computational load of the proposed receiver is mainly dictated by the generalized right-inverse operations executed in Steps 3 and 4 of Algorithm \ref{alg:alg1}, which corresponds to applying \eqref{hest} and \eqref{xest} in the iterative stage, followed by the complexity associated with the computation of the truncated SVD in step 7, corresponding to the rank-one approximation stage. Accounting a cost $\mathcal{O}(I^2J)$ to calculate the pseudo-inverse of a rank-$I$ matrix $\mat{A} \in \compl^{I \times J}$, the costs of \eqref{hest} and \eqref{xest} are, respectively, $\mathcal{O}(PN^2)$ and $\mathcal{O}(PKN^2)$. This means that each iteration of the BALS stage requires $\mathcal{O}(PN^2(K + 1))$. The second stage involves computing the SVD of the rank-1 matrix $\hat{\mat{X}}$, which implies $\mathcal{O}(TN)$.

\section{Benchmark methods}\label{benchmarks}
In this section, we present two methods used as benchmarks for evaluating the proposed receiver. Both consider a semi-unitary DMA precoder: the first with transmitted information symbols (data-aided approach) and the second with only pilots (pilot-aided approach).

Supposing that the DMA beamformer could be represented as a semi-unitary matrix $\check{\mat{F}} \in \compl^{P \times N}$, we plugg it into \eqref{hest} and expand the right inverve to get $\hat{\mat{H}} = \mat{Y}_{(1)}(\check{\mat{F}} \diamond \mat{X})^\ast\bigl[(\check{\mat{F}} \diamond \mat{X})^\trans(\check{\mat{F}} \diamond \mat{X})^\ast\bigr]^{-1}$. By using Khatri-Rao the property $(\mat{A} \krp \mat{B})^\trans(\mat{A} \krp \mat{B})^\ast = (\mat{A}^\trans\mat{A}^\ast) \odot (\mat{B}^\trans\mat{B}^\ast)$, we get $\hat{\mat{H}} = \mat{Y}_{(1)}(\check{\mat{F}} \diamond \mat{X})^\ast\bigl[(\check{\mat{F}}^\trans\check{\mat{F}}^\ast) \odot (\mat{X}^\trans\mat{X}^\ast)\bigr]^{-1}$, or $\hat{\mat{H}} = \mat{Y}_{(1)}(\check{\mat{F}} \diamond \mat{X})^\ast\bigl[(\check{\mat{F}}^\trans\check{\mat{F}}^\ast) \odot (\vet{m}\vet{s}^\trans\vet{s}^\ast\vet{m}^\herm)\bigr]^{-1}$. Since $\check{\mat{F}}^\trans\check{\mat{F}}^\ast = P\mat{I}_N$ and $\vet{s}^\trans\vet{s}^\ast = \|\vet{s}\|_\text{F}^2$, we obtain $\hat{\mat{H}} = (1/P\|\vet{s}\|_\text{F}^2)\mat{Y}_{(1)}(\check{\mat{F}} \krp \mat{X})^\ast\bigl(\mat{I}_{N} \odot \vet{m}\vet{m}^\herm\bigr)^{-1}$. The rank-one matrix constructed by the outer product $\vet{m}\vet{m}^\herm$ contains the squared magnitudes of the $\vet{m}$ entries in its main diagonal, and the Hadamard product $\mat{I}_{N} \odot \vet{m}\vet{m}^\herm$ isolates exactly those terms. Therefore, we obtain
\begin{equation}\label{hestfdft}
    \hat{\mat{H}} = (1/P\|\vet{s}\|_\text{F}^2)\mat{Y}_{(1)}(\check{\mat{F}} \diamond \mat{X})^\ast\diagof{\tilde{\vet{m}}},
\end{equation}
where $\tilde{\vet{m}} \eqdef \left[1/|m_1|^2,\ldots,1/|m_N|^2\right] \in \compl^{N \times 1}$. Using the same reasoning to \eqref{xest}, we have $\hat{\mat{X}} = \mat{Y}_{(2)}(\check{\mat{F}} \diamond \mat{H})^\ast\bigl[(\check{\mat{F}} \diamond \mat{H})^\trans(\check{\mat{F}} \diamond \mat{H})^\ast\bigr]^{-1}$ or, equivalently, $\hat{\mat{X}} = \mat{Y}_{(2)}(\check{\mat{F}} \diamond \mat{H})^\ast\bigl[P\mat{I}_{N} \odot (\mat{H}^\trans\mat{H}^\ast)\bigr]^{-1}$. As the product $\mat{H}^\trans\mat{H}^\ast$ contains the squared norms of the columns of $\mat{H}$, we have
\begin{equation}\label{xestfdft}
	\hat{\mat{X}} = (1/P)\mat{Y}_{(2)}(\check{\mat{F}} \diamond \mat{H})^\ast\diagof{\tilde{\vet{h}}},
\end{equation}
where $\tilde{\vet{h}} \eqdef \bigl[1/\fronorm{\mat{H}_{\cdot 1}}^2,\ldots,1/\fronorm{\mat{H}_{\cdot N}}^2\bigr] \in \compl^{N \times 1}$.

In addition to supposing a semi-unitary DMA precoder, we could also consider that the transmitted symbols are only pilots. In this case, we use the vector $\check{\vet{s}}$ in place of $\vet{s}$. We consider the pilots have form $\check{s}_t = e^{j(t-1)\omega}$, where we assume $\omega = 1/T$. As a result, $\check{\vet{s}} = [1,e^{j/T},e^{j2/T},\ldots,e^{j(T-1)/T}]$ is a Vandermonde vector. Hence, $\|\check{\vet{s}}\|_\text{F}^2 = T$, and \eqref{hestfdft} reduces to 
\begin{equation}
    \hat{\mat{H}} = (1/PT)\mat{Y}_{(1)}(\check{\mat{F}} \diamond \check{\vet{s}}\vet{m}^\trans)^\ast\diagof{\vet{\zeta}}.
\end{equation}
By replacing the estimated rank-one matrix $\hat{\mat{X}}$ by its construction in the pilot-aided approach and considering the transpose of \eqref{xestfdft}, we have $\hat{\vet{m}}\check{\vet{s}}^\trans = (1/P)\diagof{\tilde{\vet{h}}}(\check{\mat{F}} \diamond \mat{H})^\herm\mat{Y}_{(2)}^\trans$. Thus, we can isolate $\hat{\vet{m}}$ doing
\begin{equation}
    \hat{\vet{m}} = (1/PT)\diagof{\tilde{\vet{h}}}(\check{\mat{F}} \diamond \mat{H})^\herm\mat{Y}_{(2)}^\trans\check{\vet{s}}^\ast,
\end{equation}
for which we applied a filtering on $\hat{\mat{X}}^\trans$, i.e., $\hat{\vet{m}} \!=\! (1/T)\hat{\mat{X}}^\trans\check{\vet{s}}^\ast$.

\section{Simulation results}\label{sec:results}
In this section, we present simulation results to assess the performance of the proposed semi-blind CE method, including joint symbol and CE performance, as well as related computational burden. We consider 10$^4$ Monte Carlo runs using the parameter set $\{K,N,P,T\} = \{8,16,32,10\}$, and the transmit symbols are generated from a 64-QAM constellation.

\begin{figure}[!t]
	\centering
	\includegraphics[width=0.37\textwidth]{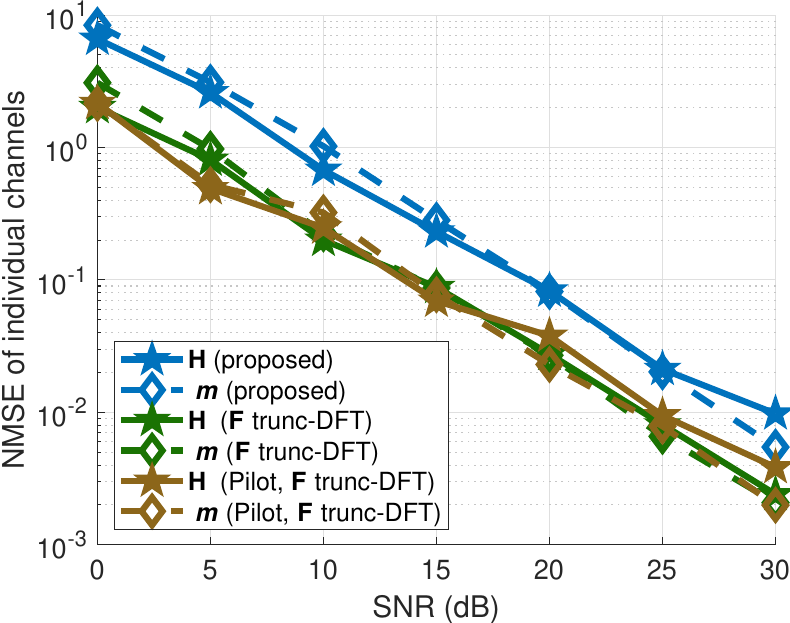}
    \caption{NMSE of individual channels Vs. SNR.}
	\label{fig:nmse}
\end{figure}

\begin{figure}[!t]
	\centering
	\includegraphics[width=0.37\textwidth]{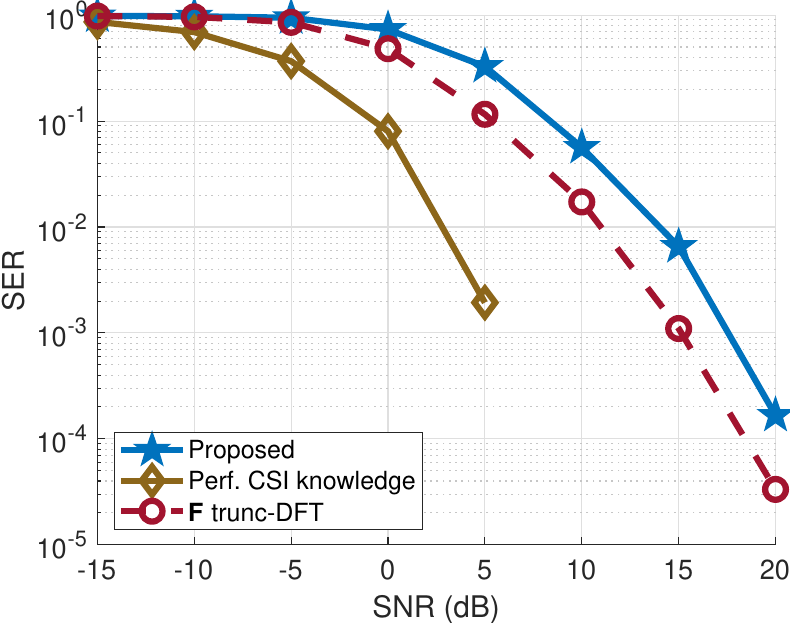}
    \caption{Symbol error rate Vs. SNR.}
	\label{fig:ser}
\end{figure}

\begin{figure}[!t]
	\centering
	\includegraphics[width=0.4\textwidth]{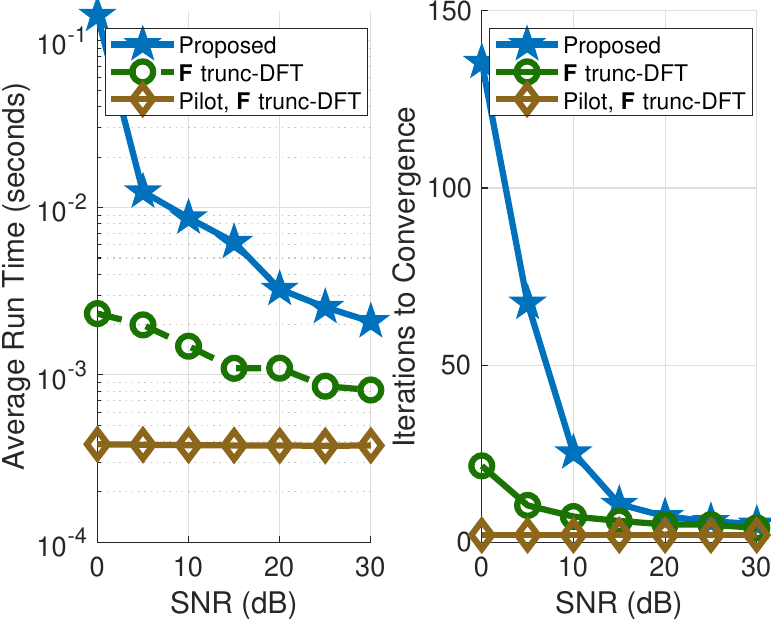}
    \caption{Computational effort in terms of average runtime and number of required iterations to ALS convergence Vs.SNR.}
	\label{fig:timeit}
\end{figure}

Figures \ref{fig:nmse} and \ref{fig:ser} show, respectively, results of normalized mean square error (NMSE) of the individual channel components $\mat{H}$ and $\vet{m}$, and symbol error rate (SER) considering the proposed two-stage method and adopted benchmark methods. As detailed in Section \ref{benchmarks}, where the first benchmark method considers the DMA precoder matrix as the truncated DFT $\check{\mat{F}}$ instead of enforcing the Lorenztian-phase constraint, while the second is somewhat more conservative, in that, in addition to accounting $\check{\mat{F}}$ it also assumes that all transmitted symbols are pilots (included in $\check{\vet{s}}$). The estimation accuracy for both $\hat{\mat{H}}$ and $\hat{\mat{m}}$ is found to be almost similar. Even though $\hat{\vet{m}}$ is estimated after two stages, and hence under the error propagation from the first to the second stage (the second one involves a the rank-1 approximation problem using $\hat{\mat{X}}$), the number of estimated channel coefficients in $\hat{\mat{H}}$ is higher, making the NMSE curves close to each other. This behavior is also observed for the benchmark methods. When a semi-unitary DMA precoder matrix takes place, a clear improvement in channel estimation performance can be observed. In fact, by comparing the NMSE curves in Fig. \ref{fig:nmse}, we noticed a gap of approximately 5 dB. In our experiments, we noticed that replacing information symbols with pilots did not provide a significant improvement in results when $\check{\mat{F}}$ was already being used. The results of symbol estimation were evaluated during the channel estimation time window due to the data-aided CE approach, with SER curves in Fig. \ref{fig:ser} following the same trend as those found for CE. Figure \ref{fig:timeit} shows that the number of iterations required for ALS convergence is similar to that of the benchmark method in the medium-to-high SNR regime. 

The overall complexity of the first stage of the Algorithm \ref{alg:alg1} depends on the SNR, while the second one is invariant to it. As a result, the BALS procedure for the proposed approach may require several iterations to converge under a low SNR regime. However, the algorithm converges fast for medium to high SNR regimes (SNR $\geq$ 15 dB). The computational cost of the proposed method can be further reduced if the DMA precoder is optimally designed. This is evident in the fast convergence of the BALS procedure when $\check{\mat{F}}$, resulting in low runtime. Considering the use of $\check{\mat{F}}$ and pilots, the computational costs are further reduced, since the receiver only needs to estimate two quantities, namely, $\mat{H}$ and $\vet{m}$, and no longer three with $\vet{s}$. Furthermore, only one matched filtering is used, avoiding matrix inversions.


\section{Conclusion}
We proposed a data-aided channel estimation framework for DMA-based systems. In the proposed approach, we have assumed that the DMA propagation vector may be unknown for generalized reasons, contrary to the literature. By capitalizing on the multilinear structure of the received signals, we derived a tensor-based semi-blind iterative receiver that estimates three quantities, namely, the wireless channel, the DMA propagation vector, and the transmitted information symbols jointly. Thanks to the data-aided approach, the proposed method can reduce symbol decoding delay by estimating and decoding symbols in advance, during the channel estimation stage. In addition, since the symbols are replicated across different subcarriers, no symbol retransmission in the time domain is needed. Estimation results under the Lorentzian-phase constraint show competitive performance with the benchmark, which accounts for DMA precoding as a semi-unitary matrix. Reducing the dependency on the number of DMA measurements while accounting for the geometric channel structure and a multi-user scenario represents one of the perspectives for continuing this research. A more challenging scenario where the DMA training matrix is not perfectly known at the receiver due to impairments, such as limited feedback, represents another problem left for future work, and involves the estimation/refinement of the DMA precoder by fully exploiting the trilinear structure of the PARAFAC tensor.

\bibliographystyle{IEEEtran}
\bibliography{references}

\end{document}